\documentclass{aa}  
\usepackage{graphicx}
\begin{document}
   \title{Determination of the basic parameters of the dwarf nova \\ EY Cygni}

   \author{J. Echevarr{\'\i}a
        \inst{1}
        \and
        R. Michel
        \inst{2}
        \and
        R. Costero
        \inst{1}
        \and
        S. Zharikov
        \inst{2}
  }

   \offprints{J. Echevarr{\'\i}a}

    \institute{Instituto de Astronom{\'\i}a, Universidad Nacional Aut\'onoma de M\'exico,
             Apartado Postal 70-264, 04510 M\'exico, D.F.,  M\'exico\\
             \email{jer,costero@astroscu.unam.mx}
         \and
             Instituto de Astronom{\'\i}a, Universidad Nacional Aut\'onoma de M\'exico,
               Apartado Postal 877, 22830, Ensenada, Baja California, M\'exico\\
             \email{rmm,zhar@astrosen.unam.mx}
             }

   \date{Received March 28, 2006; accepted ...}


  \abstract
   {High-dispersion spectroscopy of EY~Cyg obtained from data spanning
twelve years show, for the first time, the radial velocity curves from 
both emission and absorption line systems, yielding semi-amplitudes 
$K_{em}=24 \pm 4 \, \mathrm {km \, s^{-1}}$ and
$K_{abs}=54 \pm 2 \, \mathrm {km \, s^{-1}}$. The orbital period of this system
is found to be 0.4593249(1)~d. The masses of the stars, their mass ratio and 
their separation are found to be 
$M_1 \sin^3 i = 0.015 \pm 0.002 \, M_{\odot}$,
$M_2 \sin^3 i = 0.007 \pm 0.002 \, M_{\odot}$, 
$q = {K_1/K_2} = {M_2/M_1} = {0.44 \pm 0.02}$ and
$a \sin i = 0.71 \pm 0.04 \, R_{\odot}$.
We also found that the spectral type of the secondary star is around K0,
consistent with an early determination by Kraft (1962).}
   {From the spectral type of the secondary star and simple 
comparisons with single main sequence stars, we conclude that the radius 
of the secondary star is about 30 per cent larger than a main sequence 
star of the same mass.}
   {We also present VRI CCD photometric observations, some of them simultaneous 
with the spectroscopic runs. The photometric data shows several light 
modulations, including a sinusoidal behaviour with twice the frequency of the 
orbital period, characteristic of the modulation coming from an elongated, 
irradiated secondary star. Low and high states during quiescence 
are also detected and discussed.}
   {From several constrains, we obtain tight limits for the inclination angle of 
the binary system between 13 and 15 degrees, with a best value of 14 degrees 
obtained from the sinusoidal light curve analysis.} 
{From the above results we derive masses $M_1 = 1.10 \pm 0.09 \, 
M_{\odot}$, $M_2 = 0.49 \pm 0.09 \, M_{\odot}$, and a binary separation 
$a = 2.9 \pm 0.1 \, R_{\odot}$.}


   \maketitle
 
%

\section{Introduction}

Dwarf novae are a subclass of cataclysmic variables which have a semidetached 
late-type secondary star undergoing mass transfer onto a white dwarf primary. 
Outbursts are frequent in  many of these objects, with timescales ranging from a few 
days to several weeks (see review by Warner \cite{war95}). The spectra of 
dwarf novae at quiescence usually show strong and broad emission lines of H 
and He\,I, while during outburst the same lines are seen in shallow 
absorption.

EY~Cyg is a U Geminorum-type dwarf nova with quiescent V magnitude 
$\sim$14.8 and outburst V magnitude $\sim$11. Its outbursts have been reported
to have a periodicity of about 240~d (Kholopov \& Efremov \cite{ke76}) and  
durations of about 30~d (Piening \cite{pie78}), but a recent analysis 
of long term AAVSO data  may indicate that the system has gone regular outbursts 
on a 2000~d odd cycle (Tovmassian {\it et al.} \cite{tea02}). 

EY~Cyg has been detected by {\it ROSAT} in soft X-rays. Orio \& \"Ogelman (\cite{Ori92})
report a flux of $3\times 10^{-13}$ erg cm$^{-2}$ s$^{-1}$ in the 0.4-2.2 keV range,
and Richman (\cite{rich96}) finds a flux of $1.19\times10^{-12}$ erg cm$^{-2}$ s$^{-1}$ in the 
0.1-2.4 keV range. Anomalous ultraviolet flux line ratios have been 
detected by G\"ansicke {\it et al.} (\cite{gea03}) with {\it HST/STIS}; as 
the authors point out, this may indicate that the system is going through a 
phase of thermal timescale mass transfer, accreting CNO processed material 
from the external layers of the secondary star. Recently, 
Sion {\it et al.} (\cite{sea04}) derived an effective temperature of 24,000~K 
for the primary star, by fitting a white dwarf model to FUSE and STIS spectra.

Str\"omgren photometry was obtained at minimum light by Echevarr{\'\i}a, 
Costero \& Michel (\cite{ecm93}), with values $y=14.79$ and $b-y=0.53$. The 
colour index indicates a bright secondary of spectral type K, and suggests a 
very long orbital period system (Echevarr{\'\i}a, \& Jones \cite{ej84}). 2MASS 
infrared observations compiled by Hoard {\it et al.} (\cite{hea02}) give JHK 
indices for EY~Cyg compatible with those of a KO-K2 star (see their Fig. 4), 
assuming no interstellar reddening is present. First attempts to obtain a
light curve were done by Hacke \& Andronov (\cite{ha88}) from photographic 
observations, and later by Sarna, Pych \& Smith (\cite{sps95}) from R and I 
photometric observations.

Optical spectroscopy dates back to the early observations by 
Kraft (\cite{kra62}), who pointed out that EY~Cyg may have a small velocity 
variation (close to his detection limit), indicating a system viewed nearly 
{\it pole-on}. The presence of Balmer emission lines and of absorption 
lines like Fe~I $\lambda 4045$ and Ca~I $ \lambda 4226$, left no doubt as to 
the binary nature of the system. Kraft classified the secondary as a 
K0V star. Later, Szkody, Pich\'e \& Feinswog (\cite{spf90}) obtained further 
spectrograms, 12 and 13~days after an outburst, midway between maximum and 
quiescence. They reported that the spectra revealed strong and narrow emission 
lines while the absorption lines were still evident. The measured H$\gamma$ and 
H$\beta$ emission lines yielded no radial velocity curve and no indication as 
to the spectral type of the secondary star was given. 
Smith~{\it et al.}~(\cite{sscj97}), from an ISIS spectrum, reported rather weak 
Balmer and He~I emission, a strong red continuum, many absorption
features from a red dwarf component, weak TiO bands and rather weak CN lines. 
No measurements of the emission lines were given and the authors gave a
spectral classification of the secondary in the range K5-M0.

A first, tentative orbital period determination of this system was made by 
Hacke \& Andronov (\cite{ha88}), obtaining a period of 0.181228~d. Later 
Sarna, Pych \& Smith (\cite{sps95}), reported a longer orbital period with 
probable values of 0.2630 or 0.2185~d. The authors argued in favour of the 
latter value by claiming a dM2-dM3 spectral type of the secondary based on 
the spectroscopy by Smith~{\it et al.}~(\cite{sscj97}). Preliminary reports of 
a much larger orbital period (around 0.45 days) were given by Costero, 
Echevarr{\'\i}a \& Pineda (\cite{cep98}), Tovmassian {\it et al.} (\cite{tea02}) 
and Echevarr{\'\i}a {\it et al.} (\cite{eea03}). This larger period is supported
by the rate of decline from outburst observed by Kato, Uemura \& Ishioka 
(\cite{kui02}).

In this paper we present and discuss low and high dispersion spectroscopic 
observations of EY~Cyg spanning twelve years, as well as CCD VRI 
photometric measurements for this object.

\section{Observations} 

Our first observations of EY~Cyg were obtained with a Boller \& Chivens 
(B\&Ch) spectrograph attached to the 2.1m Telescope of the Observatorio 
Astron\'omico Nacional at San Pedro M\'artir, during a run from 1993 May 9 to 11,
using a 600 lines mm$^{-1}$ grating and a 19~$\mu$m Thomson 1024$\times$1024 
CCD, to cover a spectral range from 
$\lambda 5820 \, \mathrm{\AA}$ to $\lambda$7500~{\AA} with an spectral
resolution of around 3.3~\AA. 70 spectra were obtained with 300~s exposure
time for each. 

Further observations were obtained with the 
Echelle spectrograph attached to the same telescope. On the nights of 
1993 September 25 and 26 the Thomson 1024$\times$1024 CCD was used to cover 
a spectral range from $\lambda$4080 to $\lambda$6640~\AA, with spectral resolution 
R=18,000. On the subsequent runs (1998 July 13 to 17; 1999 July 22 to 24; 
1999 October 3 and 4; 2000 August 22; and 2001 August 28) the Thompson 
2048$\times$2048 detector was used to obtain a spectral resolution 
R=22,000, covering the range $\lambda$3850 -- $\lambda$7600~\AA. On 2004 July 22 and 2004 October 9 
the 1024$\times$1024 SITe3 detector was used at a spectral resolution of 
R=16,000, covering the range from $\lambda$3800 to $\lambda$6900~\AA. All these observations we carried
out with the 300~l/mm cross-dispersor, which has a blaze angle at around 5500~\AA. Our last run was conducted 
on 2005 June 26 to July 1 with the Thompson 2048$\times$2048 detector and the 150 l/mm Echellette grating
which has a blaze angle around 8000~\AA. Exposure times where 900~s for most 
observations, except a few for which 600 or 1200~s was selected. 

VRI photometry was simultaneously conducted during the three nights of the 
July 1999 spectroscopic run. This was done with the 1.5~m telescope and a 
Tektronix 1024$\times$1024~CCD. Exposure times for the V, R and I filters were 
typically 60, 30 and 25~s, respectively. An IRVVRI continuous sequence was 
conducted. Further I-filter photometry was carried out with the same telescope and 
a SITe detector on the nights of 2001 August 7, 8 and 30, and V photometry 
on the nights of 2001 August 28, 29. During the nights of 2004 July 22, 26, 
28 and 2005 June 20, 23 and 24, additional I-filter measurements were done with the 0.84~m telescope. 
More VRI photometry was conducted simultaneously to the spectroscopic observations on
2005 June 25 through July 1, and on July 18, with the same SITe detector on the 1.5~m telescope. 

The data was reduced using standard IRAF routines for CCD photometry. Secondary photometric 
stars, reported in Misselt's~(1996) catalogue, were used; the differential 
magnitudes reported in this paper are referred to the star identified as 
number 9 (V=14.630, R=15.069) in that catalogue, which was similar in 
brightness to EY~Cyg during all of our observing runs. 

A summary of all the observations is presented in Table~\ref{tab:ObsLog}. 
Several late-type spectral standard stars, including 61~Cyg A and B, were also
observed (mostly during the EY Cygni runs); their published spectral types and radial 
velocities are listed in Table~\ref{tab:VelStan}. 

\begin{table*}[!]
  \setlength{\tabcolsep}{1.0em} 
  \begin{center}
    \caption{Observing Log.}
    \label{tab:ObsLog}
    \begin{tabular}{lcccccc}
       \hline
       \noalign{\smallskip}
Date & Starting HJD & Instr./Tel. & Range(\AA)/Band & Resolution/Exp. Time & 
No. of spectra/images \\
       \noalign{\smallskip}
       \hline
       \hline
       \noalign{\smallskip}
{\bf Spectroscopy} & &&&& \\
1993 May 9-11       & 2449117.6 & B\&Ch/2.1m    & 5820-7500 &  2,000 & 70 \\
1993 Sep 25,26      & 2449255.7 & Echelle/2.1m  & 4080-6640 & 18,000 & 11 \\
1998 Jul 13-19      & 2451008.0 & Echelle/2.1m  & 3850-7600 & 22,000 & 49 \\
1999 Jul 22,23,24   & 2451381.8 & Echelle/2.1m  & 3850-7600 & 22,000 & 39 \\
1999 Oct 3,4        & 2451454.6 & Echelle/2.1m  & 3850-7600 & 22,000 & 17 \\
2000 Aug 22         & 2451778.8 & Echelle/2.1m  & 3850-7600 & 22,000 &  3 \\
2001 Aug 28         & 2452149.7 & Echelle/2.1m  & 3850-7600 & 22,000 &  9 \\
2004 Jul 22         & 2453208.9 & Echelle/2.1m  & 3800-6900 & 16,000 &  3 \\
2004 Oct 9          & 2453287.7 & Echelle/2.1m  & 3800-6900 & 16,000 &  3 \\
2005 June 26-July 1 & 2453547.9 & Echelle/2.1m  & 3800-6900 & 22,000 &  54 \\
       \noalign{\smallskip}
       \noalign{\smallskip}
{\bf Photometry} &&&&& \\
1999 Jul 22,23,24   & 2451381.7 & Direct/1.5m   & VRI    &  60/30/25   & 283/288/287 \\
2001 Aug 7,8        & 2452128.8 & Direct/1.5m   & I/I    &  180/180    & 61/98\\
2001 Aug 28,29,30   & 2452149.6 & Direct/1.5m   & V/V/I  &  40/40/20   & 480/447/815 \\
2004 Jul 22,26,28   & 2453208.7 & Direct/0.84m  & I/I/I  &  30/30/30   & 545/366/197 \\
2005 Jun 20,23,24   & 2453541.9 & Direct/0.84m  & I/I/I  &  30/30/30   & 155/193/278 \\
2005 Jun 25-Jul 1   & 2453546.7 & Direct/1.5m   & VRI    &  40/40/40   & 1016/1019/1020 \\
2005 Jul 18         & 2453569.9 & Direct/1.5m   & VRI    &  40/40/40   & 161/154/160 \\

      \noalign{\smallskip}
      \hline
    \end{tabular}
  \end{center}
\end{table*}

\section{Spectroscopic results}

The spectra of EY~Cygni show a very narrow H$\alpha$ emission line, as well as 
a very weak (sometimes absent) H$\beta$ line. Also conspicuous is the 
combination of the HeI $\lambda 5875$ line in emission and the NaI doublet in 
absorption around $\lambda 5890$. Many other absorption features are also visible, mainly those due to 
neutral Calcium, Chromium and Iron.

\subsection{Radial Velocities}

The spectra obtained with the B\&Ch spectrograph show a coherent 
radial velocity curve, although we were unable to obtain reliable 
results from them due to two factors: 
1) the semi-amplitudes of both components have 
indeed very low values; 
2) the object is in the direction of an extended emission nebula, which 
makes H$\alpha$ particularly difficult to deconvolve. We have therefore
set aside these spectra only for a discussion on the spectral type of the 
secondary, and use the Echelle observations to obtain the radial velocity 
curves from the emission and absorption components, as well as the orbital 
parameters of the binary.
Heliocentric corrections have been applied to all
calculated radial velocities.

\begin{table}
\setlength{\tabcolsep}{1.em} 
\begin{center}
  \caption{Observed Spectral Type Comparison Stars}
  \label{tab:VelStan}
      \begin{tabular}{rrr}
      \hline
      \noalign{\smallskip}
 Name      & Radial Velocity     & Spectral Type  \\
           & ($\mathrm{km \, s^{-1}}$)              &       \\
      \noalign{\smallskip}
      \hline
      \hline
      \noalign{\smallskip}
{\bf Low Dispersion} & \\
 HD175541   &  18.8     &   G8 V    \\
 HD221354   & -22.1     &   K0 V    \\
 HD190404   &  -2.6     &   K1 V    \\
 HD190007   & -28.9     &   K4 V    \\
 HD151288   & -31.0     &   K7.5 Ve \\
 HD147379A  & -20.7     &   M0 Vvar \\
 HD147379B  & -24.0     &   M4      \\
      \noalign{\smallskip}
      \noalign{\smallskip}
{\bf High Dispersion} \\
 HR8544, HR8545  &  -3.3, -6.2  &   G1 V + G2 V \\
 HR8883          &  16.1        &   G4 V        \\
 HR8734          & -14.0        &   G8 IV       \\
 $\sigma$ Dra    &  26.7        &   K0 V        \\
 36 Oph B        &   0.0        &   K2 V        \\
 HR751           &  23.4        &   K3 V        \\
 61 Cyg A        & -64.3        &   K5 V        \\
 61 Cyg B        & -64.5        &   K7 V        \\
      \noalign{\smallskip}
      \hline
      \end{tabular}
    \end{center}
\end{table}

\subsubsection{Absorption}

We have measured the radial velocity of the absorption line system of EY~Cyg by cross-correlating 
its spectra with that of 61 Cyg A independently at two dispersion orders covering the spectral intervals 
$\lambda\lambda$5200-5400~\AA and $\lambda\lambda$5350-5550~\AA, hereinafter identified as 
orders 42 and 41, respectively. In these 
orders strong Fe I lines -- like $\lambda\lambda$5232, 5269, 5270, 5282, 5283, 5324, 5328, 
5415, 5434, 5429, 5434, 5447, 5455 and 5477 \AA -- are present. These and other absorption lines 
in those orders are used simultaneously with the cross-correlation technique to obtain a radial velocity
in each spectrum, with a typical error of about 5 $\mathrm{km \, s^{-1}}$. From our
188 Echelle spectra, 21 were discarded due to their very poor signal to noise ratio.
The results derived for the remaining 167 spectra are shown in 
Tables~\ref{tab:RadVelSep93}~through~\ref{tab:RadVel05}.

\subsubsection{Emission}

EY~Cygni is in the field of a diffuse, complex emission nebulosity (Tovmassian {\it et al.} \cite{tea02}; 
Sion {\it et al}. \cite{sea04}). This complicates the process of extracting 
radial velocities from the emission lines. Furthermore, visual inspection of
our original CCD Echelle images shows that the nebular H$\alpha$ emission
is not symmetric at both sides of the object's spectrum. We have therefore
applied a careful cleaning procedure to the spectra, in order to subtract 
the nebular emission before deriving the radial velocities. 

\begin{figure}
  \begin{center}
     \includegraphics[width=0.8\columnwidth,angle=90]{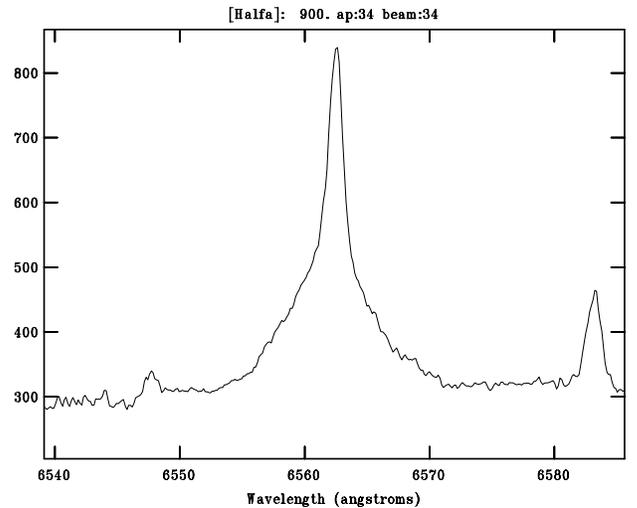}
     \caption{The added spectra at the H$\alpha$ emission line region.}
     \label{fig:Halfa}
  \end{center}
\end{figure}

To illustrate 
this problem, we show in Figure \ref{fig:Halfa} the added spectrum around the H$\alpha$ region,
using all the available spectra. Since the nebular components were not subtracted,
the [NII] emission lines are clearly present, as well as the narrow and 
stationary nebular H$\alpha$ (FWHM $\sim1 \,$\AA) superimposed on a broader 
(FWHM $\sim5 \,$\AA) emission line, blurred here by the radial velocity shifts 
of the disc along all orbital phases. The broad component is slightly blue-shifted 
with respect to the nebular emission due to the systemic velocity of the binary 
(see Table~\ref{tab:OrbitalParameters} below). 

The "cleaning" of  the nebular 
component was done by fitting, in each individual spectrum, a Gaussian to the 
narrow component, and then subtracting it from the original spectrum. This procedure 
gave better results than using sky subtracted spectra. We also visually inspected 
the spectra obtained by both methods, to check that no residuals of the nebular 
emission were left on the broader line. 

\begin{figure}
  \begin{center}
     \includegraphics[width=0.8\columnwidth]{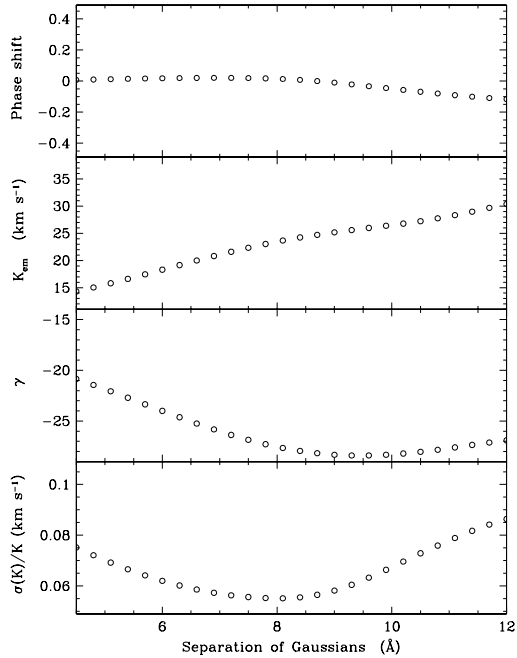}
     \caption{Diagnostic diagram for EY~Cyg. The best estimate of the 
semi-amplitude of the white dwarf is 28~km~s$^{-1}$, corresponding to  
s=8.1\AA.}
     \label{fig:diagnos}
  \end{center}
\end{figure}

The H$\alpha$ emission line was measured using the standard double Gaussian 
technique and the diagnostic diagrams described by Shafter, Szkody \& 
Thorstensen (1986), to whom we refer for the details on the interpretation of 
this diagnostic tool. We have used the {\it convolve} routine from the IRAF 
{\it rvsao} package, kindly made available to us by J.R. Thorstensen. The emission line 
is at times so weak that, in the case of 34 spectra, we were unable to obtain any 
measurement at all. H$\beta$ was too weak to derive any reasonable results from it. 
Due to the heterogeneity of our different spectroscopic runs, the "cleaned" 
H$\alpha$ spectra were first re-binned to a standard linear dispersion of 
$0.15\,$ \AA\  per pixel, a value corresponding to the mean spectral resolution 
of all runs in that Echelle order. The double-Gaussian program was 
then run with a FWHM value of $4.5 \, $\AA for the individual Gaussians, 
using a wide range of values for $s$, the separation between Gaussians. 
The results are shown in Figure~\ref{fig:diagnos}.

\begin{figure}
  \begin{center}
    \includegraphics[width=\columnwidth]{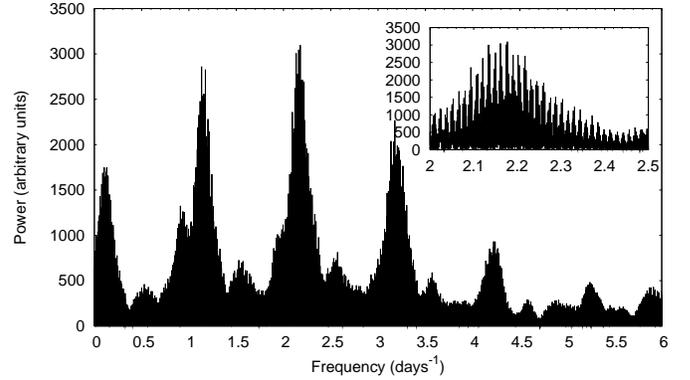}
    \caption{Periodogram for the absorption radial 
             velocities of all observing runs.} 
    \label{fig:periodogram_abs}
  \end{center}
\end{figure}

\begin{figure}
  \begin{center}
    \includegraphics[width=1.0\columnwidth]{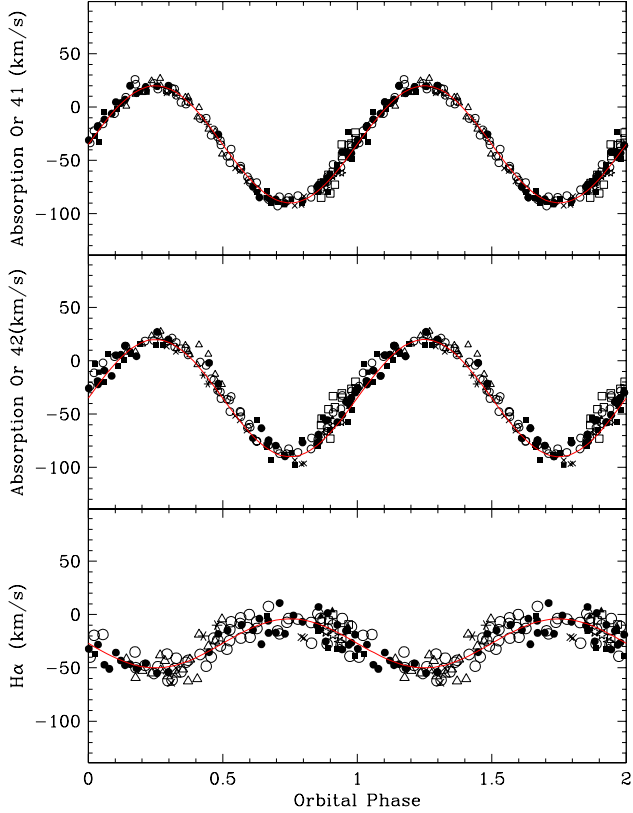}
    \caption{Folded (0.4593249~d) radial velocities for all the observing runs. 
             Top: Absorption lines in Order 41;
             Middle: Absorption lines in Order 42; and 
             Bottom: H$\alpha$ emission radial velocities.
             The symbols correspond to:
             open squares, 1993 September;
             solid squares, 1998 July;
             solid circles, 1999 July;
             open triangles, 1999 October;
             solid triangles, 2000 August;
             crosses, 2001 August; 
             asterisks, 2004 July;
             stars, 2004 October; and
             open circles, June 2005.} 
    \label{fig:Specorb}
  \end{center}
\end{figure}

\begin{figure}[t]
  \begin{center}
    \includegraphics[angle=270,width=\columnwidth]{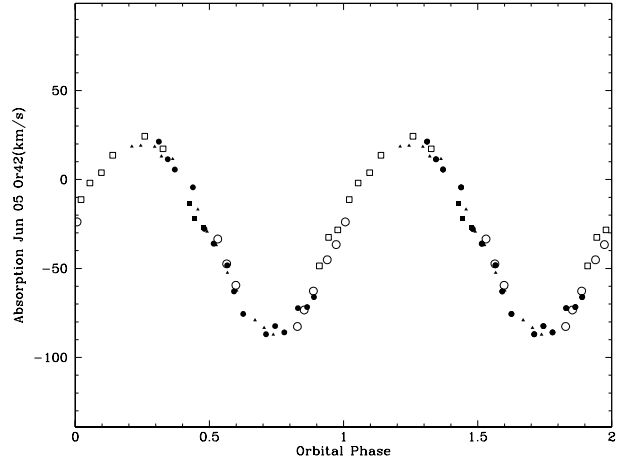}
    \caption{Radial velocity curve for the Junes 2005 run.
             The symbols correspond to:
             solid triangles, June 27;
             dots, June 28;
             open circles, June 29;
             open squares, June 30:
             filled squares, July 1.
             } 
    \label{fig:junevel}
  \end{center}
\end{figure}

As expected from the presence of an asymmetric low-velocity component, the 
semi-amplitude should asymptotically approach the correct value as $s$ 
increases. On the other hand, there is a change in slope in the $\sigma(K)/K$ 
curve when $s$ has reached the line width at the continuum 
and thereafter the velocity measurements become dominated by noise. 
The systemic velocity decreases rapidly, up to a value of 
$s \approx\,8\,$\AA, and remains flat afterwards. 
On the other hand, $K_{em}$ vary very 
slowly and is the parameter least affected by the different solutions. The best radial 
velocity results, which are shown in the second column of Tables \ref{tab:RadVelSep93} through 
\ref{tab:RadVel05}, are obtained for $s = 8.1$\AA. 
   
\begin{table}[!]
  \setlength{\tabcolsep}{1.1em} 
  \begin{center}
    \caption{Radial velocities for 1993 September 25-26.}
    \label{tab:RadVelSep93}
    \begin{tabular}{rrrr}
       \hline
       \noalign{\smallskip}
HJD        & H$\alpha$   &  Absorption &  Absorption \\
           &             &  Order 41   & Order 42 \\
(2400000+)  & ($\mathrm{km \, s^{-1}}$)      &  ($\mathrm{km \, s^{-1}}$)     \\
       \noalign{\smallskip}
       \hline
       \noalign{\smallskip}
49255.72620 &   -6.6  &  -71.1 & -54.1 \\
49255.73800 &   -5.4  &  -63.5 & -45.4 \\
49255.74981 &   -0.0  &  -60.9 & -33.2 \\
49255.76439 &  -22.6  &  -45.9 & -34.4 \\
49255.77619 &         &  -34.4 & -26.2 \\
49255.78800 &         &  -29.6 & -26.4 \\
49256.64421 &         &  -85.0 & -60.2 \\
49256.65601 &         &  -81.1 & -73.1 \\
49256.66920 &         &  -72.8 & -43.5 \\
49256.68032 &  -14.9  &  -35.0 & -30.6 \\
49256.69143 &   -9.3  &  -36.0 & -38.2 \\
49256.70254 &         &  -24.0 & -19.6 \\
      \noalign{\smallskip}
      \hline
    \end{tabular}
  \end{center}
\end{table}
 
\begin{table}[!]
  \setlength{\tabcolsep}{1.0em} 
  \begin{center}
    \caption{Radial velocities for  1998 July 13-17.}
    \label{tab:RadVelJul98}
       \begin{tabular}{rrrr}
          \hline
          \noalign{\smallskip}
HJD        & H$\alpha$   &  Absorption &  Absorption \\
           &             &  Order 41   & Order 42 \\
(2400000+)  & ($\mathrm{km \, s^{-1}}$)      &  ($\mathrm{km \, s^{-1}}$)     \\
          \noalign{\smallskip}
          \hline
          \noalign{\smallskip}
51007.95822 &  -1.4 &  -78.5 &        \\
51007.96988 &  24.1 &  -90.3 &  -93.0 \\
51008.86086 &       &  -80.0 &  -55.2 \\
51008.87449 &       &  -84.9 &  -79.1 \\
51008.89246 &       &  -88.7 &        \\
51008.90626 &       &  -91.5 &  -88.2 \\
51008.91954 &       &  -86.5 &        \\
51009.88395 & -15.5 &  -80.1 &  -79.7 \\
51009.89866 & -31.9 &  -80.0 &        \\    
51009.91316 & -31.9 &  -54.2 &        \\
51009.93304 & -15.6 &  -23.0 &  -57.3 \\
51009.94836 & -39.3 &  -35.8 &  -27.3 \\
51009.96277 & -36.5 &  -27.5 &   -3.4 \\
51010.77446 &       &  -90.1 &  -74.1 \\
51010.80816 &       &  -69.4 &  -53.2 \\
51010.82181 &       &  -63.0 &  -64.8 \\
51010.89214 &       &   -4.2 &  -19.6 \\
51010.91763 &       &        &   -4.5 \\
51011.68127 &       &        &  -97.7 \\
51011.77820 &       &  -48.4 &  -46.5 \\
51011.80284 &       &  -32.8 &  -10.7 \\
51011.81606 &       &        &    6.1 \\
51011.84696 &       &        &    0.9 \\
51011.86113 &       &        &    5.8 \\
51011.87423 &       &   15.3 &   15.5 \\
51011.88799 &       &   14.1 &        \\
51011.90224 &       &        &   15.3 \\
51011.91574 &       &        &   14.5 \\
       \noalign{\smallskip}
       \hline
    \end{tabular}
  \end{center}
\end{table}

\begin{table}[!]
  \setlength{\tabcolsep}{1.0em} 
  \begin{center}
     \caption{Radial velocities for 1999 July 22-24.}
     \label{tab:RadVelJul99}
     \begin{tabular}{rrrr}
       \hline
       \noalign{\smallskip}
HJD        & H$\alpha$   &  Absorption &  Absorption \\
           &             &  Order 41   & Order 42 \\
(2400000+)  & ($\mathrm{km \, s^{-1}}$)      &  ($\mathrm{km \, s^{-1}}$)     \\
       \noalign{\smallskip}
       \hline
       \noalign{\smallskip}
51381.80533 &  -9.6 &       &       \\
51381.82489 & -14.6 & -74.8 & -72.5 \\
51381.86300 & -17.0 & -86.7 & -79.6 \\
51381.87993 & -18.0 & -90.1 & -89.6 \\
51381.93263 & -10.2 & -74.7 & -77.8 \\
51381.97642 & -32.7 & -42.2 & -50.8 \\
51381.99335 & -26.3 & -43.4 & -34.8 \\
51382.66710 & -35.2 &       &  -2.0 \\
51382.68447 & -18.1 & -25.7 & -21.5 \\
51382.70115 & -14.8 & -24.5 &       \\
51382.75382 &  -3.8 & -84.7 &       \\
51382.77101 & -17.2 & -85.4 & -74.4 \\
51382.85283 & -14.3 & -72.5 & -75.7 \\
51382.86968 & -25.2 & -70.5 & -55.7 \\
51382.88635 &  -9.6 & -59.8 & -57.6 \\
51382.90660 & -28.4 & -47.1 & -40.2 \\
51382.92465 & -32.2 & -31.1 & -25.7 \\
51382.94167 & -28.2 & -17.7 & -22.1 \\
51382.96195 & -51.0 &  -6.2 & -14.2 \\
51382.97880 & -43.1 &   3.7 &   5.6 \\
51382.99564 & -46.6 &  19.9 &   8.5 \\
51383.67261 & -27.9 &       & -63.2 \\
51383.68947 &  -5.9 &       & -76.4 \\
51383.70741 &  10.8 &       &       \\
51383.73015 &  -0.9 &       &       \\
51383.74674 &  -6.7 & -87.8 & -76.6 \\
51383.77345 &   7.1 & -74.3 & -69.1 \\
51383.79019 &   1.5 & -66.3 & -55.3 \\
51383.81866 & -15.1 & -42.3 & -38.5 \\
51383.83534 & -18.9 & -35.9 & -29.8 \\
51383.85230 &       & -18.8 & -18.9 \\
51383.86894 & -47.1 & -11.7 &  -9.1 \\
51383.88916 & -35.7 &   4.6 &   5.2 \\
51383.90584 & -47.1 &   6.9 &  14.2 \\
51383.92257 & -50.3 &  13.0 &   4.3 \\
51383.93931 & -45.8 &  18.9 &       \\
51383.96084 & -54.9 &  19.5 &   27.1\\
51383.97815 & -53.9 &  20.0 &   19.8\\
      \noalign{\smallskip}
      \hline
    \end{tabular}
  \end{center}
\end{table}

\begin{table}[!]
  \setlength{\tabcolsep}{1.0em} 
  \begin{center}
    \caption[]{Radial velocities for 1999 October 3-4, 2000 August 22, 
               2001 August 28, 2004 July 22 and 2004 October 9.}
    \label{tab:RadVel99-04}
    \begin{tabular}{rrrr}
      \hline
      \noalign{\smallskip}
HJD        & H$\alpha$   &  Absorption &  Absorption \\
           &             &  Order 41   & Order 42 \\
(2400000+)  & ($\mathrm{km \, s^{-1}}$)      &  ($\mathrm{km \, s^{-1}}$)     \\
      \noalign{\smallskip}
      \hline
      \noalign{\smallskip}
51454.63882 & -41.9 &    0.9 &   8.0 \\
51454.65322 & -59.4 &   12.0 &   8.7 \\
51454.66637 & -52.6 &   13.6 &  15.8 \\
51454.68405 & -55.0 &   24.7 &  23.6 \\
51454.69796 & -52.3 &   20.1 &  27.6 \\
51454.71122 & -57.8 &   19.5 &  19.5 \\
51454.72886 & -56.2 &   12.8 &  11.6 \\
51454.74755 & -62.5 &    9.4 &   8.6 \\
51454.76506 & -60.3 &   -9.3 &  -3.7 \\
51455.61267 & -38.2 &   26.8 &  22.0 \\
51455.62595 & -43.5 &   14.7 &  17.2 \\
51455.65597 & -51.3 &    9.2 &  14.4 \\
51455.66984 & -42.1 &    3.6 &   5.7 \\
51455.68395 & -18.7 &   -0.4 &  15.0 \\
51455.69785 & -29.9 &  -20.7 &   6.2 \\
51455.71167 & -25.1 &  -23.1 & -12.7 \\
51455.72476 &  -4.3 &  -44.2 & -22.2 \\

51778.79744 &  -1.6 &  -70.3 & -70.6 \\
51778.81398 &   3.9 &  -66.1 & -53.4 \\
51778.84496 & -11.4 &  -40.7 & -43.7 \\

52149.66676 & -52.4 &   12.9 &  13.4 \\
52149.67523 & -64.0 &   14.1 &  17.8 \\
52149.68299 & -50.0 &   10.8 &   8.4 \\
52149.78664 &       &  -57.4 & -51.6 \\
52149.79432 &       &  -59.9 & -56.0 \\
52149.80199 &       &  -69.8 & -65.4 \\
52149.88871 &       &  -92.5 & -93.3 \\
52149.89639 & -21.2 &  -91.7 & -96.8 \\
52149.90407 & -21.9 &  -83.8 & -96.3 \\

53208.93353 & -20.1 &  -12.0 & -13.6 \\
53208.94475 &       &  -16.1 & -21.7 \\
53208.95631 &  -9.9 &  -26.0 & -26.8 \\

53287.68969 & -16.1 &  -75.5 & -66.7 \\
53287.70084 & -18.6 &  -64.8 & -60.7 \\ 
53287.71199 & -25.5 &  -61.6 & -57.8 \\
      \noalign{\smallskip}
      \hline
    \end{tabular}
  \end{center}
\end{table}

\begin{table}[!]
  \setlength{\tabcolsep}{1.0em} 
  \begin{center}
    \caption[]{Radial velocities for 2005 June 26 - July 1}
    \label{tab:RadVel05}
    \begin{tabular}{rrrr}
      \hline
      \noalign{\smallskip}
HJD        & H$\alpha$   &  Absorption &  Absorption  \\
           &             &  Order 41   & Order 42 \\
(2400000+)  & ($\mathrm{km \, s^{-1}}$)      &  ($\mathrm{km \, s^{-1}}$)     \\
      \noalign{\smallskip}
      \hline
      \noalign{\smallskip}
53547.91137 &  -37.4 & -10.6 &  -1.1 \\
53547.93209 &  -22.1 & -21.4 & -17.9 \\
53547.94793 &  -12.4 & -35.5 & -35.4 \\
53547.96377 &  -14.1 & -52.3 & -47.7 \\ 
53548.71897 &  -49.8 &  21.4 &       \\
53548.73562 &  -47.7 &  15.6 &  18.6 \\
53548.75147 &  -55.2 &  22.1 &  19.0 \\
53548.77325 &  -62.0 &  16.7 &  18.4 \\
53548.78910 &  -50.7 &  16.9 &  13.0 \\
53548.80494 &  -44.8 &   2.8 &  11.6 \\
53548.84648 &  -39.8 & -26.5 & -16.8 \\
53548.86233 &  -29.5 & -31.5 & -29.3 \\
53548.87817 &  -24.6 & -46.2 & -36.8 \\
53548.89810 &  -22.2 & -58.2 & -52.4 \\
53548.91395 &  -21.3 & -70.8 & -62.4 \\
53548.92979 &  -13.1 & -77.6 & -75.3 \\
53548.94857 &    7.6 & -86.7 & -79.1 \\
53548.96441 &  -15.3 & -92.8 & -83.4 \\
53548.98026 &   -7.3 & -88.0 & -87.1 \\
53549.69693 &  -61.4 &  17.6 &  21.3 \\
53549.71278 &  -53.7 &   4.4 &  11.4 \\
53549.72862 &  -42.9 &   5.3 &   5.6 \\
53549.76122 &  -48.4 & -10.5 &  -4.4 \\
53549.77707 &  -24.1 & -28.0 & -27.7 \\
53549.79291 &  -32.5 & -46.2 & -36.0 \\
53549.81408 &   -3.0 & -55.4 & -48.2 \\
53549.82993 &  -12.2 & -65.4 & -62.9 \\
53549.84578 &   -6.8 & -70.8 & -75.6 \\
53549.88380 &  -18.2 & -84.0 & -87.0 \\
53549.89964 &   -3.7 & -84.6 & -82.4 \\
53549.91549 &   -4.2 & -83.1 & -85.9 \\
53549.93616 &  -26.8 & -77.4 & -72.3 \\
53549.95201 &   -3.6 & -77.2 & -71.7 \\
53549.96785 &  -13.6 & -66.2 & -66.1 \\
53550.71858 &   -6.7 & -39.5 & -33.5 \\
53550.73443 &   -8.3 & -57.3 & -47.4 \\
53550.75028 &   -0.0 & -68.4 & -59.5 \\
53550.85211 &   -7.5 & -84.6 & -82.6 \\
53550.86795 &    0.7 & -78.0 & -73.3 \\
53550.88380 &  -27.5 & -70.4 & -62.7 \\
53550.90879 &  -28.6 & -56.6 & -45.1 \\
53550.92463 &  -39.0 & -41.4 & -36.6 \\
53550.94049 &  -40.1 & -33.7 & -23.8 \\
53552.73156 &   -9.7 & -53.9 & -48.5 \\
53552.74756 &   -3.7 & -48.5 & -32.5 \\
53552.76341 &  -11.2 & -37.0 & -28.3 \\
53552.78316 &  -19.5 & -22.3 & -11.3 \\
53552.79900 &  -18.6 & -15.9 &  -2.0 \\
53552.81485 &        &  -6.2 &   3.8 \\
53552.83491 &        &   9.3 &  13.6 \\
53552.85076 &  -35.1 &  26.0 &       \\
53552.89061 &  -48.9 &  15.8 &  24.3 \\
53552.90648 &  -37.0 &  14.9 &       \\
53552.92232 &  -40.9 &  11.2 &  17.2 \\

      \noalign{\smallskip}
      \hline
    \end{tabular}
  \end{center}
\end{table}

\subsection{The Orbital Parameters and Radial Velocity Curve}

To determine a tentative orbital period we have first made a thorough period 
search on the absorption radial velocity data, by running a periodgram power 
spectrum (Deeming~1975). The results are shown in Figure~\ref{fig:periodogram_abs} (top), for 
which the optimum frequency value is 2.177105(80)~d$^{-1}$, equivalent to a
a preliminary period of 0.459333(2)~d. On the lower part of the Figure, we show a close-up of the
power spectrum region between 2 and 2.5~~d$^{-1}$ to check for possible aliases between 0.4 and 0.5 days.
Based on this period and assuming
initial values for the systemic velocity, the semi-amplitudes and the zero 
point, we then proceeded to obtain the orbital parameters by fitting the 
measured radial velocities, of both the absorption and the emission data, with 
sinusoids of the form 

$$V(t) = \gamma + K \sin[2\pi (t - \mathrm{HJD_o})/P_{orb}]\,,$$

based on a least-squared algorithm. The solutions to the absorption component
are very similar for both orders, so we have adopted their average. The results for
the absorption and the emission systems are listed in Table \ref{tab:OrbitalParameters}. 
The measured radial velocities, together with their sinusoidal fits, are 
plotted in Figure \ref{fig:Specorb}. The data has been folded with an orbital period of 0.4593249~d 
derived from the solution for the absorption lines, a value which we will adopt hereinafter
as the orbital period of the binary. The zero phase correspond to the inferior 
conjunction of the secondary star. We must point out that the finding of this orbital period value
is strongly supported by the fact that during the June 2005 run, we obtained radial velocity
data during four consecutive nights and an additional subsequent night (see Table~\ref{tab:RadVel05}),
that rule out an alias period, as shown in Figure~\ref{fig:junevel}. This is further supported by the
photometry of the same four consecutive nights simultaneous to the spectroscopic run, shown in 
section 4.3 (see Figure~\ref{fig:low05I}).

The radial velocity curves in Figure~\ref{fig:Specorb} show better 
results for the secondary star than for the emission line. This is not 
surprising, considering that the semi-amplitude of the former is almost twice than
that derived from H$\alpha$, and the above mentioned complications in obtaining a nebular-free 
disc emission. All considered, the systemic velocities and orbital periods of 
both components agree within the errors,
although it is possible that we are underestimating the 
semi-amplitude of the white dwarf as we will discuss below.
 
\begin{table}
 \begin{center}
   \caption{Spectroscopic Orbital Parameters of EY~Cyg.}
   \label{tab:OrbitalParameters}
   \begin{tabular}{@{}ccr@{}}
      \hline
Orbital   & Absorption &{H$\alpha$ emission} \\
Parameter &      &     \\
      \noalign{\smallskip}
      \hline
      \noalign{\smallskip}
$\gamma$ (km s$^{-1}$)          & -33 $\pm$ 2  & -28 $\pm$ 4   \\
$K$ (km s$^{-1}$)               & \,54 $\pm$ 2 & \,24 $\pm$ 4  \\
$\mathrm{HJD_o}$ (2449255 +)          & 0.3260(9)    & 0.337(5)      \\
 $P_{\mathrm{orb}}$ (days)          & 0.4593249(1) & 0.459323(3)   \\
${\sigma}$                      &  6.8         & 10.3          \\
     \noalign{\smallskip}
     \hline
   \end{tabular}
 \end{center}
\end{table}

Therefore, throughout this paper, we adopt the ephemeris:

$$\mathrm{HJD} = 2,449,255.3260(9) \, + \, 0.4593249(1) \,E \, ,$$

from the inferior conjunction of the secondary star.

\subsection{Spectral Type of the Secondary Star}

\begin{figure*}[t]
  \begin{center}
     \includegraphics[width=\columnwidth]{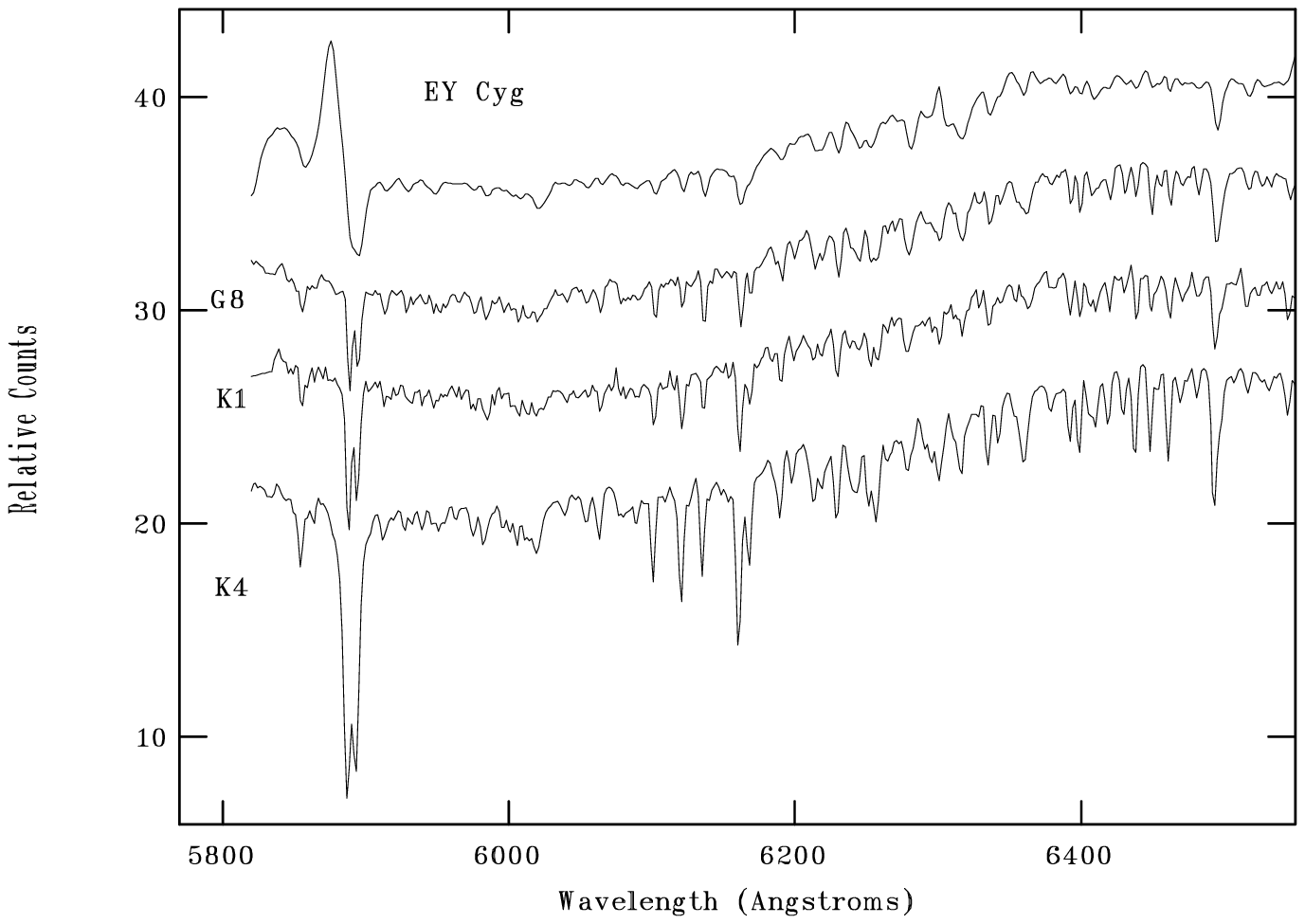}
     \includegraphics[width=\columnwidth]{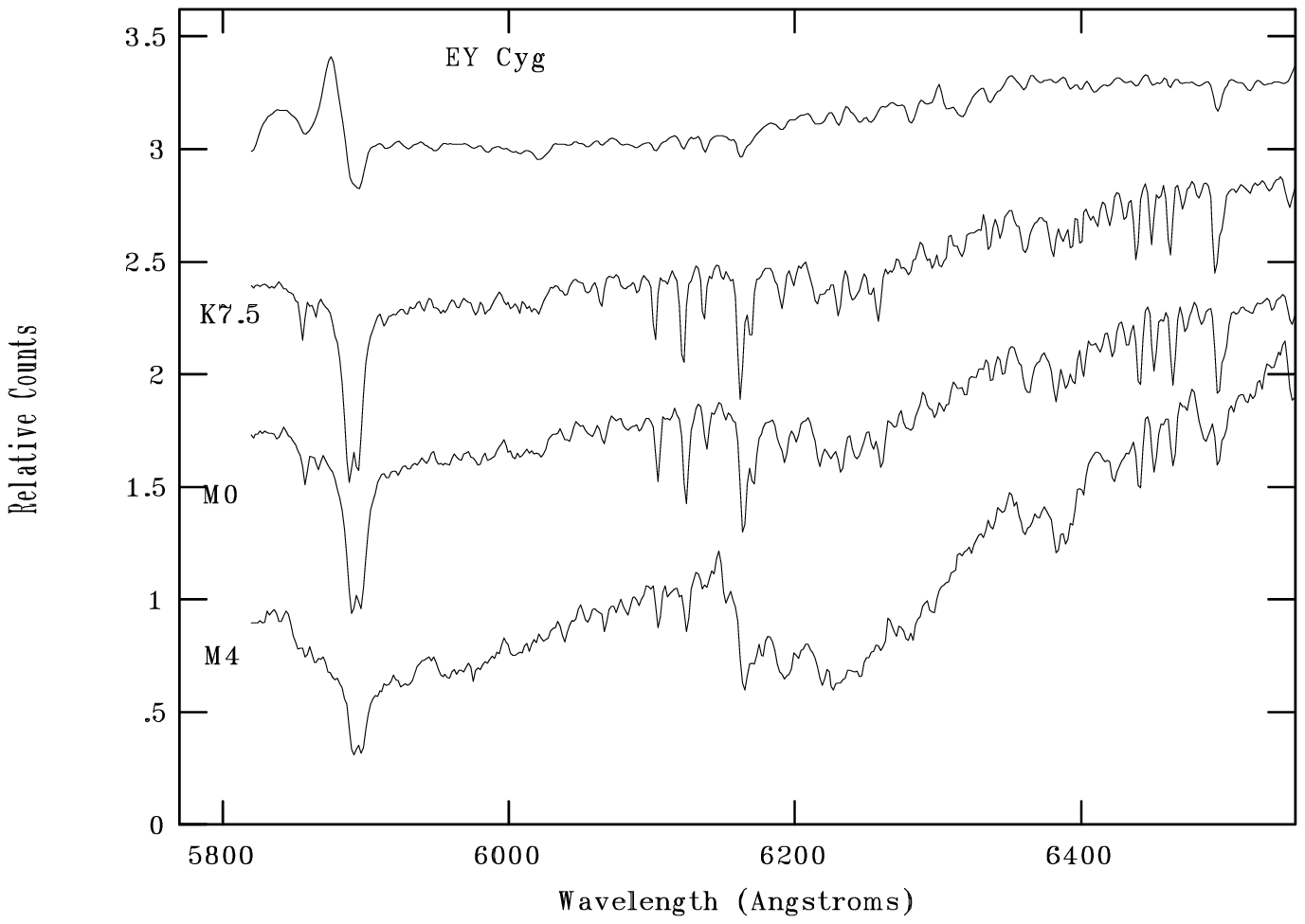}
     \caption{The mean, low-dispersion spectrum of EY~Cyg compared with G8- through M4-type stars.}
     \label{fig:fig5}
  \end{center}
\end{figure*}

A low dispersion mean spectrum of EY~Cyg has been constructed by co-adding the individual 
B\&Ch spectra in the frame of reference of the secondary star, using the 
orbital parameters in Table~\ref{tab:OrbitalParameters}. The resultant  
"absorption enhanced", or {\it co-added} spectrum is shown at the top of 
Figure \ref{fig:fig5}. At the left, the object is compared
with spectra of main sequence stars classified in the range
G8 to K4, while on the right, a comparison is made with stars
of later spectral types within the range K7.5 to M4. Details of these observed spectral comparison stars,
obtained with the same setup as the B\&Ch spectra are included in Table~\ref{tab:VelStan}. 

Based on a spectrum obtained by Smith {\it et al.}~(1997), Sarna, Pych, \& 
Smith~(1995) argued that the secondary star corresponds to a spectral type 
dM2-dM3. However, our {\it co-added} EY Cyg spectrum do not correspond to this classification. 
No traces of TiO bands are present in our spectra, as shown in Figure \ref{fig:fig5}. A comparison of these spectra 
with our object yields to the conclusion that the spectral type of 
the secondary star of EY~Cyg is that of a late G or early K star. 
To refine the spectral type we compare the
high resolution spectra of the object with a number of spectral type standard stars in the range 
G8 to K5 (see Table~\ref{tab:VelStan}).

\begin{figure}
  \begin{center}
    \includegraphics[width=1.\columnwidth]{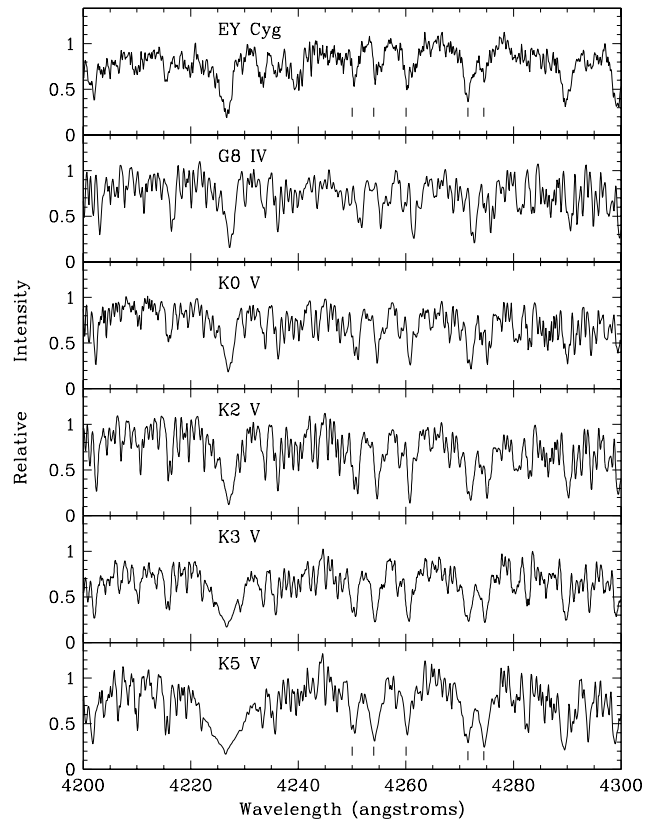}
    \caption{Co-added Echelle spectrum of EY~Cyg in the $\lambda\lambda$4200-4300~\AA\ region,
             compared with several standard spectral-type stars (see text).}
    \label{fig:fig6}
  \end{center}
\end{figure}

Since the Echelle spectra are separated in different orders and our observational
setup was not intended to add the orders and flux calibrate the overall spectra, we are not able to present similar 
results as in the low resolution spectra. Instead, we have looked for specific spectral signatures in several orders. 
The results are shown here for one particularly important spectral region, 
for which we have {\it co-added} a mean spectrum of EY~Cyg, in the same manner as in the case of the low 
resolution spectra. Figure \ref{fig:fig6} shows the region $\lambda\lambda$4200-4300~\AA. A 60 percent 
flat continuum was subtracted to the spectrum of EY Cyg and rescaled, in order to compensate for the continuum 
arising from the other light sources in the system.
The best lines for spectral classification in stars, which 
are independent of chemical abundances are the line ratios of the FeI lines 
$\lambda 4250$, $\lambda 4260$, $\lambda 4271$ \AA~ to
the CrI lines $\lambda 4254$ and $\lambda 4274$ \AA, as well as the strength of the 
CaI $\lambda$4226~\AA\ (Keenan \& McNeil 1976; Yamashita {\it et al.} 1978). 
The intensity of the CaI line increase steadily as the spectral type advances (in fact this line 
has a saturation effect for late K an M stars). On the other hand
the CrI lines also increases
in strength with respect to the FeI lines with spectral type. These FeI and CrI lines have
been marked at the top and bottom of the Figure for clarity purposes. 
Both the strength of the CaI line and the CrI/FeI ratios of EY Cyg are compatible with a
spectral type between G8 and K0.
Hereinafter, in accordance with Kraft (\cite{kra62}), we adopt a K0 spectral
type for the secondary star.

\subsection{The $M_1 - M_2$ Diagram}

To calculate the masses of the binary one may assume that the semi-amplitudes of
the measured radial velocities respresent the true orbital motion of the stars. In cataclysmic
variables, however, this may not be an accurate assumption, as the emission lines arising from
the disc may suffer several asymmetric distortions (see Wade 1985 for a full discussion on
the difficulties in measuring $K_{\mathrm{em}}$). Likewise, absorption line profiles may be subjected
to irradiation or hot-spot contaminations in the surface of the secondary, distorting therefore the
true value of $K_{\mathrm{abs}}$ (Wade \& Horne 1988). These problems were severe in the early determinations
of the radial velocities, but modern high resolution spectroscopy has greatly improved the methods in
detecting such assymmetries to provide more reliable radial velocity values (Warner 1995). 
It is important to do the best possible job on the determination of the masses of the binary, because, as mentioned
in the introduction, EY Cyg displays a very low C/N abundance ratio, which is expected to be the hallmark
of CVs that went through thermal-timescale mass transfer 
( Schenker {\it et al.} 2002; 
G\"ansicke {\it et al.} 2003). If this
is correct, one could expect that the white dwarf in EY Cyg grew in mass
during that phase - and thus confirming a relatively high white dwarf mass.
Taking these into
consideration we now attempt to estimate the masses of EY Cyg using the derive radial velocity
semi-amplitudes.

Adopting the 0.4593249~d orbital period from the absorption lines solution, 
and taking 

$$K_1=K_{em}=24 \pm 4 \, \mathrm{km \, s^{-1}}$$ and
$$K_2=K_{abs}=54 \pm 2 \, \mathrm{km \, s^{-1}},$$

we obtain:

$$M_1 \sin^3 i = {P K_2 (K_1 + K_2)^2 \over 2 \pi G} = 0.0156 \pm 0.002 M_{\odot}\,,$$
$$M_2 \sin^3 i = {P K_1 (K_1 + K_2)^2 \over 2 \pi G} = 0.0069 \pm 0.002 M_{\odot}\,;$$
the projected binary separation
$$ a \sin i = {P (K_1 + K_2) \over 2 \pi} = 0.71 \pm 0.04 R_{\odot}\,;$$
and the mass ratio 
$$q = {K_1/K_2} = {M_2/M_1} = {0.44 \pm 0.02}\,.$$

\begin{figure}
   \begin{center}
      \includegraphics[width=\columnwidth]{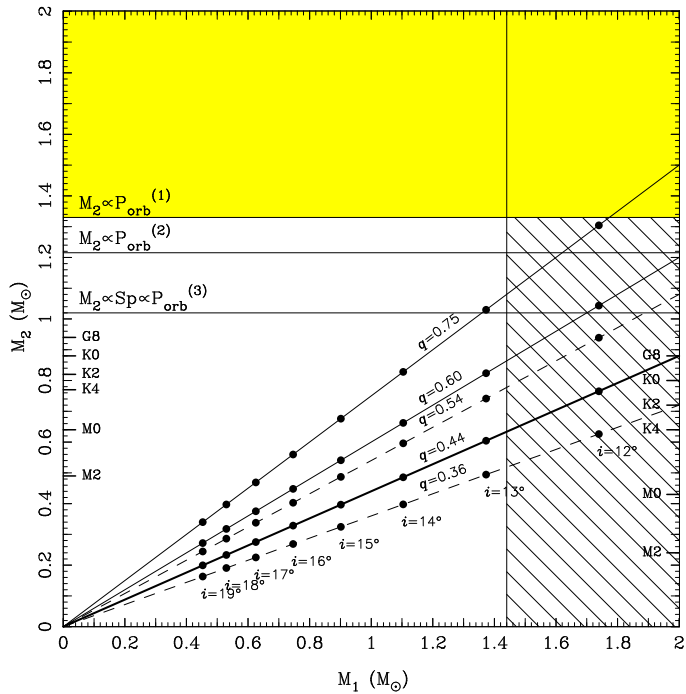}
      \caption{$M_1-M_2$ diagram. See text for explanation.}
      \label{fig:m2m1}
   \end{center}
 \end{figure}

Although EY~Cyg is a low inclination binary, for a cataclysmic variable a lower limit and an upper estimate 
of the inclination angle can be derived by using some reasonable assumptions. To illustrate this, we plot the 
solutions  for the masses of the components as a function of the mass ratio $q$ and the inclination
angle $i$. Such an $M_1-M_2$ diagram is 
shown in Figure~\ref{fig:m2m1}. The diagonal solid line correspond to the mass 
ratio, $q=0.44$, derived in this paper. The adjacent dashed lines are upper and lower 
limits of $q=0.54$ and $q=0.36$, obtained when the extreme values of the semi amplitudes are used,
instead of the error for $q$ quoted above. 
The dots along these lines correspond to different 
mass solutions for specific values of the inclination angle. The vertical 
line is at the Chandrasekhar mass limit and divides the white dwarf degenerate regime from that
of neutron stars. This mass limit could be as low as 1.2~$M_{\odot}$ for a high metalicity 
white dwarf star (Weinberg 1972). The horizontal upper region (shown in grey) is 
limited by the $M_2-P_{orb}$ relation obtained by Warner (1976) using a 
theoretical ZAMS mass-radius relation (label 1). The other two horizontal 
limits arise from the mass-radius relation by Echevarr{\'\i}a (1983) for 13 spectroscopic 
and 12 visual binaries with well detached main sequence stars (label 2),
and from the $M_2-Sp$ and $Sp-P_{orb}$ relations in Kolb \& Baraffe (2000)~(label 3). 
The spectral types labelled on the $M_2$ axis correspond to masses of ZAMS stars 
models (left) and to evolved main sequence stars (right), under some extreme assumptions as 
discussed by Kolb \& Baraffe (2000) (see upper curve in their Figure~2).
Here, we most stress
that the use of these limits and of any mass-radius-spectral type relations, should be taken with
great caution. They are used in Figure~\ref{fig:m2m1} with the goal to make the $M_1-M_2$ diagram a useful tool 
for the analysis of masses in CVs for which only $K_1$, $K_2$ and the spectral type of the secondary are known.
Secondary stars in these systems are almost certainly not main sequence stars (Echevarr{\'\i}a 1983;
Beuermann {\it et al.} 1998). However, if we want to improve on the 
M-R-Spectral Type relations used in our $M_1-M_2$ diagram, new and detailed models are needed
on evolved main sequence stars and on stars with thermal-timescale mass transfer stages 
(Shenker {\it et al.} 2002; G\"ansicke {\it et al.} 2003; Sion {\it et al.} 2004). 

From the $M_1 - M_2$ diagram and the above constrains, the inclination angle 
of the binary cannot be lower than about 13 degrees. Moreover, inclination 
angles greater than 15 degrees will yield very low masses for the secondary 
star, incompatible with the observed spectral type. In this respect, the scale 
of the spectral types, shown in the left of the diagram will move down whether we consider 
either of two mechanisms. On the one hand, if an evolved 
secondary is taken into account, then the 
spectral type for a given mass will be earlier than that of a ZAMS star 
(see Figure 2 in Kolb \& Baraffe \cite{kolb00}), as shown in the right side of the diagram.
On the other hand, if we have an X-ray heated secondary -- a possibility supported by the detection of 
soft X-rays in EY~Cyg (see the introduction) -- then the relative radius of the 
secondary, as well as its luminosity and temperature, will increase 
(Hameury {\it et al.} 1993) and, consequently, the spectral type will 
also be earlier than a ZAMS star. For our system we have no independent data that could tell us how much 
each of these two mechanisms contribute. However the observed spectral 
type, K0, suggests that both effects contribute significantly. This is evident as
the first mechanism alone will yield $i=13^o$ only for $q=0.54$. 
In section 4.2.1 we favour an inclination angle of 14 degrees, based on our
observed sinusoidal light curve.

\subsection{The radius of the secondary}

The secondary appears larger than a ZAMS star for the same 
mass. The argument is as follows: if we assume that the secondary 
is a ZAMS star then a K0 spectral type corresponds to $M_2=0.87~M_{\odot}$
(Figure 2 in Kolb \& Baraffe \cite{kolb00}). 
This corresponds to a ZAMS star radius of about 0.80~$R_{\odot}$ 
(Figure 2 in Patterson 1984). However, for a mass of 0.87~$M_{\odot}$ and 
orbital period of 11.02~h, the radius of the secondary 
is 1.13~$R_{\odot}$ (equation 4 in Echevarr{\'\i}a
1983). The difference is a factor of 1.4. 

On the other hand the mass of the secondary, as derived from the radial velocity curves and the best solution
for the inclination angle (see section 4.2.1), is $M_2 = 0.52 M_{\odot}$. In this case, the
corresponding ZAMS radius will be about 0.46 $R_{\odot}$,  but now
the radius of the secondary from Echevarr{\'\i}a's relation yields 0.96~$R_{\odot}$. The difference is a
factor of 2.1.

An increase of the radius of the secondary star by a factor between 1.4 and 2.1
can be explained through the mechanisms discussed above. In the case of an X-ray heated star, the radius 
can be increased by a factor as large as two (Hameury {\it et al.} \cite{ham93}), 
while an evolved main sequence star  scenario also implies a significantly 
larger stellar radius (Kolb \& Baraffe \cite{kolb00}) (see our cautionary statement in section 3.4).

In this respect the case of EY Cyg is similar to the long
orbital period system DX And (10.6 h), whose secondary has been found 
to have a K1V spectral type and a radius a factor of 1.4 larger than a
corresponding main sequence star (Drew, Jones \& Woods \cite{djw93}).

\subsection{The rotational velocity of the secondary}

The observed rotational velocity of the secondary star could provide an independent determination on the
mass ratio of the binary (e.g. Wade \& Horne 1988). The rotational velocity of a semi-detached tidally locked binary
is given by:

$$ V_{rot} \, \sin i = (K_{em} + K_{abs}) \, R_{RL}/a,$$

(Horne, Wade \& Szkody 1986), where $R_{RL}/a$ is the Roche-Lobe radius in units of the separation $a$. If we combine
this relation with the analytical approximation by Echevarr{\'\i}a (1983):

$$ R_{RL}/a = 0.47469 \, [q/(1+q)]^{1/3},$$

based on the tabulations by Kopal (1959) and accurate to 2 per cent for $ 0.6 < q < 1.25 $, we obtain:

$$ V_{rot} \, \sin i = 0.47469 \, K_{abs}\, q^{1/3} \, (1+q)^{2/3}.$$

\begin{figure}[h]
  \begin{center}
    \includegraphics[angle=270,width=\columnwidth]{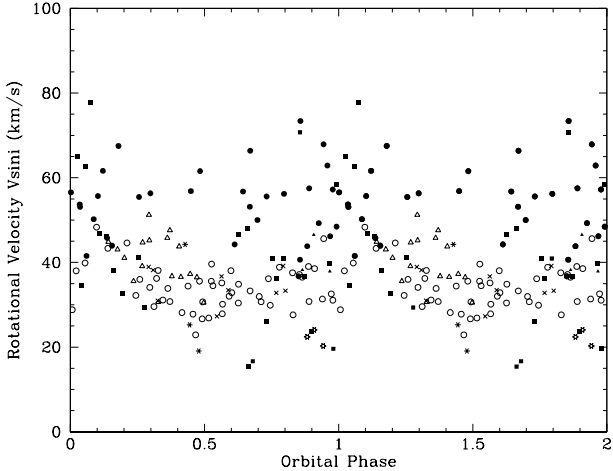}
    \caption{Rotational velocity of the secondary star.
             The symbols correspond to:
             solid squares, 1998 July;
             solid circles, 1999 July;
             open triangles, 1999 October;
             solid triangles, 2000 August;
             crosses, 2001 August; 
             asterisks, 2004 July;
             stars, 2004 October; and
             open circles, June 2005.} 
    \label{fig:rot42}
  \end{center}
\end{figure}

The width of the correlation used to calculate the radial velocities can be used also to derive an observed
rotational velocity. To convert the measured $\sigma$ to $V_{rot} \, \sin i $ we have made a calibration by 
broadening the template star 61 Cyg A, with a suitable rotational kernel for different rotational values. 
The broadened templates 
are then cross-correlated with the original template and their $\sigma$ is calculated with a gaussian fit. We have 
used a simple broadening function for spherical bodies as described by Gray (1976), using a limb
darkening coefficient of $\epsilon$ = 0.5 and a bin width of 2.23 km s$^{-1}$. A kernel has been produced 
for a range of $V \sin i$ from 10 to 200 km s$^{-1}$ using the IRAF program {\it convolve} 
to broadened the template star.

A plot of the derived values of $V_{rot} \, \sin i$ 
as a function of phase is shown in Figure~\ref{fig:rot42} for EY Cyg.
The width of the correlation was converted to rotational velocities taking into account the spectral resolution,
obtained at each observing run, which depend on the detector used and varies from 14 to 18 km s$^{-1}$. 
The Figure shows two clearly distinct values;  one with $V \sin i \, \approx $ 50 km s$^{-1}$ 
(filled dots and squares, mainly) and another with  $V \sin i \, =$ 34 $\pm$ 4 km s$^{-1}$ (open circles only). 
Using these values and the equation derived 
above we obtain $q=1.35$ and $q=0.75$ respectively. The first value is incompatible with the 
radial velocity results and also with the expected value of the secondary mass for a large orbital period system
(Echevarr{\'\i}a 1983). Furthermore, it is obtained from the observing runs related to the {\it high} state (see
next section) and its true value may be distorted by heating effects on the secondary, or they may be the result
of a large noise dispersion due to the low signal to noise ratio of the absorption lines which are
not prominent at these {\it high} 
states. On the other hand, the low rotational value, obtained during the {\it low} state (see next section) 
is more consistent with our radial velocity results, although, a mass ratio of $q=0.75$, imposes more stringent
limits in the  $M_1 - M_2$ diagram and implies a low primary mass. We have also calculated a mean value of 
30 km s$^{-1}$ for those observations near orbital phase 0.5 (superior conjuntion) which should be least 
affected by possible ellipsoidal variations. This value implies $q=0.6$ and $K_1=40$ km s$^{-1}$,
if we keep the calculated $K_2$ fixed. This value of $K_1$ is still within the maximum and minimum values shown
in the H$\alpha$ radial velocity plot in Figure~\ref{fig:Specorb}. We conclude by arguing that, for
this low inclination system, the results of the rotational velocity analysis should be taken only as a cross check
on the radial velocity results and not as an independent method of obtaining the mass ratio.

\begin{figure}[!]
  \begin{center}
    \includegraphics[width=\columnwidth]{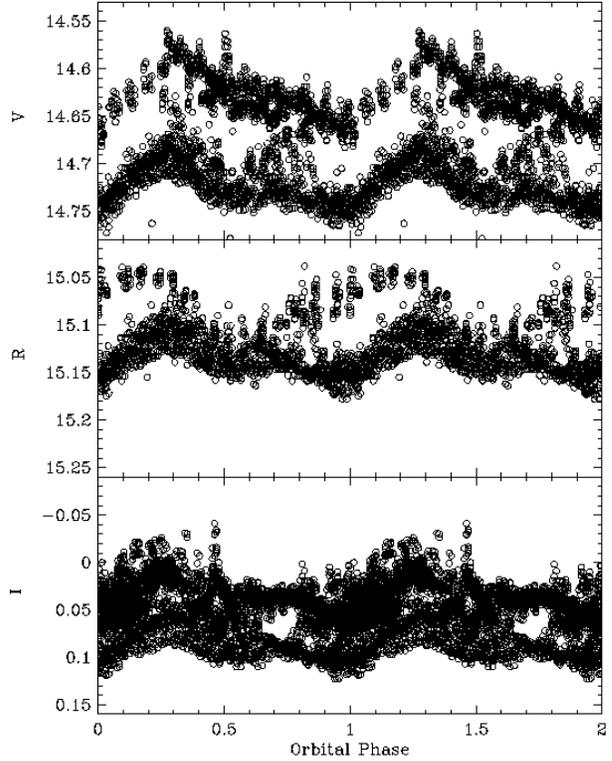}
    \caption{The complete set of our photometric observations
    folded with the adopted orbital period,
    all obtained during quiescence. A very
    complex behaviour is seen, with distinguishable {\it high} and {\it low} states (see text).}
    \label{fig:fotall}
  \end{center}
\end{figure}

\section{Photometric results}

Due to the complexity of the light curves observed during several years we have divided the discussion
in different sections.

\subsection{Overall results}

Figure~\ref{fig:fotall} shows our photometric observations of EY Cyg, all obtained during quiescence, 
folded with the adopted orbital period. These were gathered during several seasons from 
1999 to 2005 (see Table~\ref{tab:ObsLog}). The light curves show a modulation of about 
0.08 magnitudes, with a general trend to show a maximum around phase
0.25 and a minimum near phase 0.0. We find also that the average brightness of the system varies 
between seasons. This variation is clearer in V where the change is as large as 0.1 mag. During the 
fainter states, a lower envelope, reminiscent of the ellipsoidal light variations from an elongated 
secondary star is clearly seen in all filters (see section 4.3). Other features, less clear in this figure, 
appear only occasionally and are correlated with the brighter quiescent stages of the object (see section 4.2).
We will hereinafter refer to these relatively faint and bright episodes as the {\it low} and {\it
high} states. We must clarify here that this nomenclature refers only
to the different light curve behaviour seen during quiescence in this
system, and must not be compared with the quiescent and maximum stages
in Anti-Dwarf Novae or in Polar systems (Warner 1995). The outburst
behaviour in EY~Cyg has not been properly studied, as no detailed
light curves have been obtained during this stage.

\subsection{The {\it high} state}
 
\begin{figure}
  \begin{center}
    \includegraphics[width=\columnwidth]{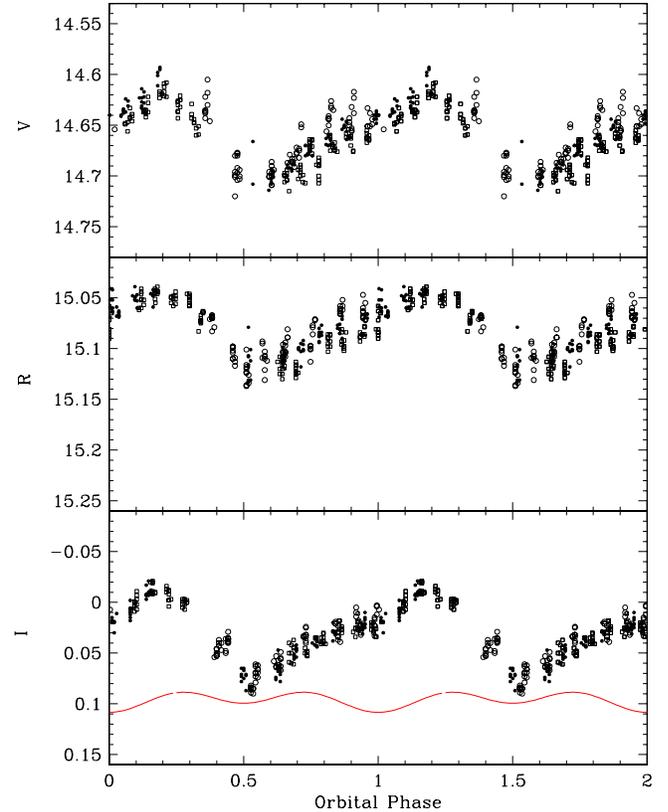}
    \caption{Folded photometric measurements for the July 1999 run.
             The symbols correspond to:
             open circles, July 22;
             solid dots, July 23; and
             open squares, July 24.
             } 
    \label{fig:fot99}
  \end{center}
\end{figure}

\begin{figure}[!]
  \begin{center}
    \includegraphics[width=\columnwidth]{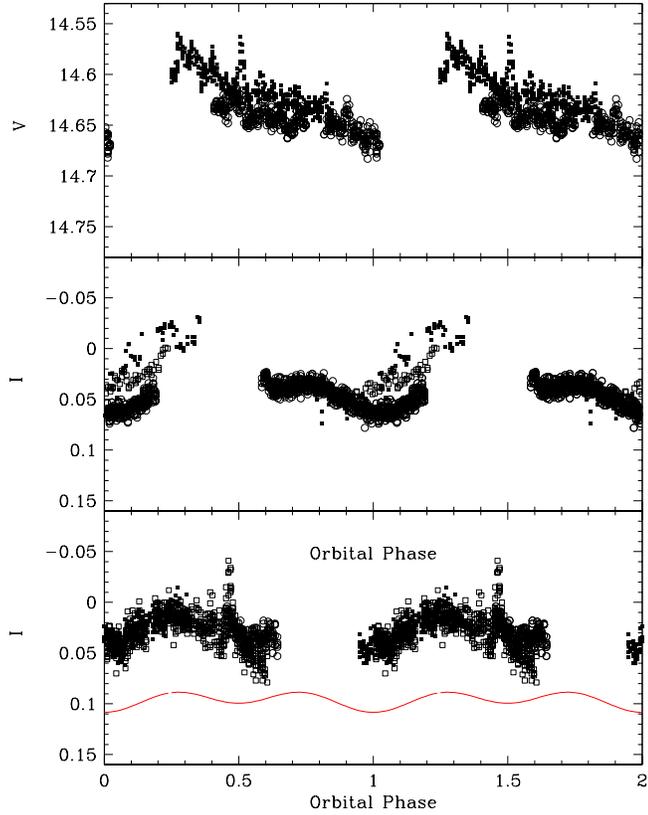}
    \caption{Non simultaneous V and I CCD photometry of the 2001 (upper and midle pannels) 
      and 2004 (lower panel) runs.    
             The symbols correspond to: 
             Filter V (2001, top);
             solid squares, August 28;
             open circles, August 29.
             Filter I (2001, middle);
             solid squares, August 07;
             open squares, August 08;
             crosses, August 30;
             Filter I (2004, bottom);
             solid squares, July 22;
             open squares, June 26;
             crosses, June 28.
            }  
    \label{fig:fot0104}
  \end{center}
\end{figure}

In the {\it high} state the light curve shows a different behaviour during our different 
observing runs, like a flat and slow increase immediately followed by a rapid decrease, as well as the 
opposite behaviour -- a rapid increase followed by a slow decrease --, but always with the maximum 
around phase 0.25 and with no counterpart around phase 0.75 (see Figures~\ref{fig:fot99} to~\ref{fig:fot05}). 
For example, Figure~\ref{fig:fot99} shows that during the 1999 observations, obtained in three contiguous nights
and covering most of the orbital phases, the light curve in all VRI filters has a minimum at orbital phase 0.5 and a
{\it slow} increase up to a maximum at phase 0.25. In contrast to this behaviour, in the V data obtained
during two consecutive nights at the end of August 2001 (upper panel in Figure~\ref{fig:fot0104} there is again a 
maximum at phase 0.25, but the brightness now decreases slowly, up to at least phase 0.0 
(in this run, there is a gap in the data between this phase and 0.25).
The geometrical location of this {\it asymmetric source} and its origin are, by no means, obvious. 
It is a puzzling result which can not be explained by a simple source. If it were a spot on the secondary, 
it would have to be located on the receding face of the star, which is unlikely, and a hot spot in an accretion disc 
is expected to be more prominent at phase 0.75. In either case, explanations of this sort are difficult
to conceive in a low inclination system like EY Cyg. Perhaps it is more likely to propose a synchronized primary with
an accretion column onto its magnetic pole, but this column would have to be about 90 degrees in longitude 
and at {\it ad hoc} latitude. Although synchronization of the primary in polars is possible 
(e.g. Lamb \& Melia 1988; Campbell 1989), studies on the longitude of the magnetic pole show a strong clustering around 
20 degrees (Cropper 1988). Nevertheless, polar systems have short orbital periods (< 4.6 hr) and, aside from the
soft X-ray emission detections in EY Cygni (see introduction) there is no further reason to believe this is a AM~Her type
object. The possibility of a case for an Intermediate Polar with a slow primary rotator (i.e. close to synchronization 
with the orbital period) is not completely ruled out by our photometric observations, as they have a limited time
coverage. These, and other possible scenarios would have to be tested against new and extensive photometric observations.

Figure~\ref{fig:fot0104} shows non simultaneous observations obtained during 2001 and 2004. We must note that in the
V observations on the upper panel, discussed above, the brightness of the object decreases slightly on the second night
(August 29) and that in the following night, observed with filter I (crosses, middle panel), we detect a completely new 
behaviour on the light curve. Instead of the flat and slow increase --or the opposite behaviour--, as that shown in 
Fig.~\ref{fig:fot99} and in the upper panel in Fig.~\ref{fig:fot0104}, we now observe a small sinusoidal modulation about
0.1 mag fainter than that seen in 1999 with the same filter. The other two nights, shown in this panel, correspond to 
earlier observations (August 7 and 8). Although scarce, they follow the {\it high} state pattern. In the lower
panel of this figure we show further I observations, obtained during three non consecutive nights at the end of July 2004.
In spite of the limited phase coverage, the light pattern is again sinusoidal-like, with an intermediate brightness, 
similar to the I observations of 2001, August 30. These observations appear to be at an intermediate stage between 
the {\it high} and {\it low} states.

\subsection{The {\it low} state and the elongated secondary}

\begin{figure}[!]
  \begin{center}
    \includegraphics[width=\columnwidth]{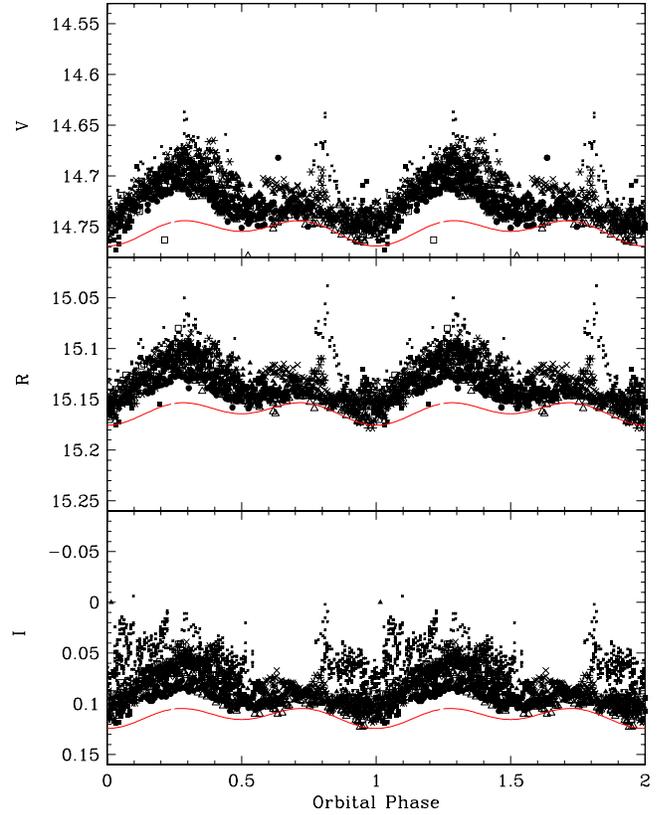}
    \caption{Photometry of EY Cyg during the 2005 run. Each symbol corresponds to
    a different night. The solid lines are the results of a model for a back illuminated
    ellipsoidal secondary, shifted downwards in the plots to facilitate their view (see text).
            }  
    \label{fig:fot05}
  \end{center}
\end{figure}

Figure~\ref{fig:fot05} shows the 2005 observations obtained while the system was at a {\it low} state,
most of which were acquired during seven consecutive nights, each of these covering from 6 to 8.2 hours,
with a mean of 7.4 hours per night. Being the orbital period 11.0238 days, these data secure a dense and well 
distributed phase coverage. Even at this {\it low} state there is still evidence of the contribution from the 
light source which shows its maximum around phase 0.25. However, the ellipsoidal modulation of the secondary 
is now evident. At the lower envelope we observe similar maxima at phases 0.25 and 0.75 -- except in filter V --, 
but different minima at phases 0.0 and 0.5, as expected from a back illuminated elongated secondary. 
The solid lines in each filter are downshifted solutions to an elongated secondary model.

\begin{figure}[!]
  \begin{center}
    \includegraphics[angle=270,width=1.0\columnwidth]{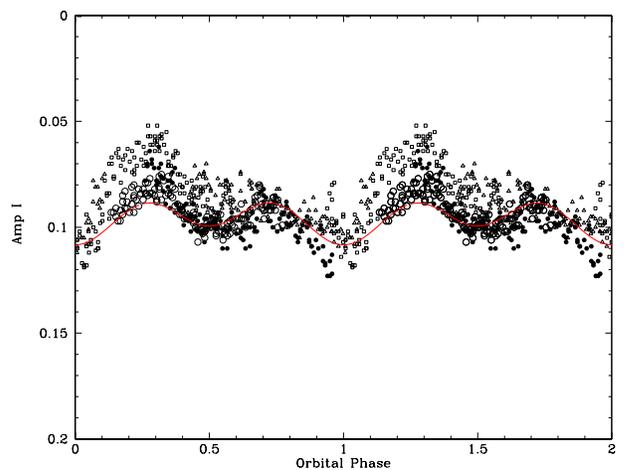}
    \caption{I band CCD photometry of selected nights during the 2005 June run.              
            The symbols correspond to:
             open squares, June 26;
             open circles, June 27;
             dots, June 28;
             open triangles, June 29.
            }  
    \label{fig:low05I}
  \end{center}
\end{figure}

We show in Fig~\ref{fig:low05I} the I filter data of the two consecutive nights of the above mentioned run, 
(large open and filled circles), during which the object showed the lowest activity from the {\it asymmetric source}.  
The other five nights in that run show a stronger activity from this source; if used, they will distort the 
shape of the light curve from the elongated secondary star (discussed in the previous section), and hence 
produce a spurious effect on the amplitude of this modulation, which is a basically an indicator of the orbital 
inclination of the binary. To show the level of contamination from this unwanted source, also plotted in the figure
are the data from the previous and following nights (small dots and
open triangles). In these, we observe that the {\it asymmetric source}
starts to dominate not only around phase 0.25, but also distorts the
minimum at phase 0.5 observed in the two quiet nights. The solid
line in this figure is the light curve modelling of the I band data
from the two quiet nights, using the program {\it Nightfall} (Wichmann
\cite{W04}), based on Roche geometry, to simulate the light curve from an elongated secondary.

The fit to the data was obtained using the following main assumptions and program options:
a) simple binary system model with a circular orbit and synchronous rotation; 
b) adopted orbital period and mass ratio calculated in section 3.2; 
c) Roche-Lobe-filling secondary star with $\mathrm T_{\mathrm eff}=4750K$ (corresponding to a spectral class K0)and 
   primary star with $\mathrm T_{\mathrm eff}=22,000K$ (Sion {\it et al.} \cite{sea04}); 
d) for the calculation of the heating effect on the secondary star, the bolometric albedo was fixed to 1.0 for the 
   primary and 0.5 for the secondary.

The I light curve fit, shown in Fig~\ref{fig:fot05}, is the same as the one shown in
Fig~\ref{fig:low05I}, but has been slightly shifted downwards for clarity purposes, whereas the
V and R light solid curves are the calculated results from the program, and are not fits to 
the corresponding observations. These calculations are in qualitatively 
good agreement with the observations. 

The best fit solution for the RV curve and the I band data is for an inclination angle of 14 degrees. With
very scarce data at a minimum state and probably still contaminated by other light sources, we cannot
claim precision to fit the light curve of an elongated secondary. Therefore
the result for the inclination angle, in particular, is taken simply as a best choice which is consistent
with the constraints derived in the $M_1 - M_2$ diagram (see section 3.4).
 
\section{Masses of the binary}

We make some final remarks on the masses of the binary, using the tight limits on
the inclination angle discussed in section 3.4. If we use the sinusoidal results from the photometric
fit, from which an inclination of 14 degrees yields the best solution, the following masses are derived:
$M_1 = 1.10 \pm 0.09 \, M_{\odot}$, $M_2 = 0.49 \pm 0.09 \, M_{\odot}$. This favours a massive primary in 
support of Costero {\it et al.} (\cite{cep98}) and Sion {\it et al.} (\cite{sea04}). The binary separation in 
this case would be $a = 2.9 \pm 0.1 \, R_{\odot}$. If we take the upper limit for the mass of the primary, set by the 
Chandrashekar limit, then the mass of the secondary is around 0.64 $M_{\odot}$, with a corresponding ZAMS spectral 
type of M0 or an evolved K4 in the extreme case discussed by Kolb \& Baraffe (2000). The latter will be more in agreement
with our observed spectral type. On the opposite extreme, if we set the inclination angle to 15 degrees, we
obtain $M_1 = 0.9 \, M_{\odot}$ and $M_2 = 0.4 \, M_{\odot}$. The spectral type of the secondary, in this case, would
have to be later than M0 in disagreement with the observations. 

\section{The distance to the system}

A rough estimate on the distance to EY Cyg can be made on the assumption the the radius of the secondary
appears to be larger than a normal K0 main sequence star. If we take our spectral standard $\sigma$ Dra as the basis
for absolute magnitude $M_V = 5.87$ (e.g. see Luck \& Heiter 2005) and assume that the secondary star contributes
50 per cent of the total flux at visual wavelengths --a reasonable extrapolation from the results in section 3.3- then
taking $m_V = 14.75$ (see Figure 13), we obtain $m_V(sec)=15.5$. The absolute magnitude for the star with radii
1.4 and 2.1 larger than the normal K0V star are $M_V = 5.14$ and $M_V = 4.36$ respectively. These values lead to
distances of 1,180 and 1,770 parsecs respectively. At these distances, the interstellar extinction should play
a distintive role. Unfortunately, the available IUE spectra (e.g. la Dous 1990) are too noisy to allow a 
determination via the $\lambda ~2200$ \AA ~dip. On the other hand, a determination using an E(B-V) excess is
unwarranted. In fact the observed (B-V)=0.73 (Misselt (1996) is bluer than that of a $\sigma$ Dra, (B-V)=0.79 or
that of the general calibration for a KOV star, (B-V)=0.82 given by Johnson (1966). This does not mean that
there is no interstellar extinction, but simply that the latter is cancelled by the blue continuum excess
that masks the true magnitude value of the secondary star. We have made above a first order estimate of the
contribution of this source at visual wavelengths. We can also make a first order estimate of
the interstellar extinction. If we assume a value A$_V$ = 1 mag/kiloparsec 
our distances calculated above are reduced to about 800 pc and 1,200 pc respectively. These
values are still larger than upper limit of 700 pc set by Sion {\it et at.} (2004), based on the X-Ray luminosity and
the background Cygnus supperbubble. We conclude that there are too many uncertainties in the case of EY Cyg to
give a reliable distance estimate. This distance determination is important in the composite accretion disc and white
dwarf analysis from the FUSE and HST observations by Sion {\it et al.} (2004). Unfortunately we are unable to favour
any of their models in particular, which are distance dependent, as our method to determine the distance to the system
does not improve previous determinations.

\section{Summary}

The most important results of our study can be summarised as follows:

1. We have been able to detect, for the first time, the radial velocity curve 
of EY~Cyg of both the emission and absorption components, their analysis 
yielding semi-amplitudes $K_{em}=24 \pm 4 \, \mathrm{km \, s^{-1}}$ and 
$K_{abs}=54 \pm 2 \, \mathrm{km \, s^{-1}}$.

2. From the analysis of the spectral data obtained during 12 years, an orbital 
period of 0.4593249~d is found.

3. We also find that the spectral type of the secondary star is near K0, 
consistent with an early determination by Kraft (1962). The secondary 
contributes around 40 per cent to the total light near 
$\lambda4300 \,$\AA, and has a radius larger than a main 
sequence star of the same mass by a factor between 1.4 and 2.1.

4. From the Chandrasekhar limit and the observed spectral type of the 
secondary, tight limits can be imposed to the inclination angle of the binary 
system, with values between 13 to 15 degrees. 

5. The light curve of the system in quiescence shows a complex behaviour, in which we
have identified {\it high} and {\it low} states. At the latter, we detect
a phase modulation with twice the frequency of the orbital phase, which can be explained by
the effect of an elongated Roche-lobe filling secondary star. During the {\it high} state
the light curve is characterized by the presence of an {\it asymmetric source} around phase 0.25
whose behaviour is different from one observing run to the other. We have discussed possible interpretations
for this phenomenon, but we conclude that new and extensive photometric observations are needed, before
advancing a convincing explanation. 

6. From the ellipsoidal fit to the lower envelope of the light curve we have derived
an inclination angle of 14 degrees, which yields masses $M_1 = 1.10 \pm 0.09 \, 
M_{\odot}$, $M_2 = 0.49 \pm 0.09 \, M_{\odot}$, and a binary separation 
$a = 2.9 \pm 0.1 \, R_{\odot}$.

\begin{acknowledgements}
    This work was partially supported by the UNAM-DGAPA project IN110002.
\end{acknowledgements}

\end{document}